# Cosmological number density in depth from $V/V_m$ distribution

Dilip G. Banhatti

School of Physics, Madurai Kamaraj University, Madurai 625021

**Abstract.** Using distribution $p(V/V_m)$ of $V/V_m$ rather than just mean $<V/V_m>$ in $V/V_m$-test leads directly to cosmological number density $n(z)$. Calculation of $n(z)$ from $p(V/V_m)$ is illustrated using best sample (of 76 quasars) available in 1981, when method was developed. This is only illustrative, sample being too small for any meaningful results.

### Luminosity-distance and volume

For cosmological populations of objects, distance is measured by (monochromatic) luminosity-distance $\ell_\nu(z)$ (at frequency $\nu$), function of redshift $z$ of object. Similarly, volume of sphere passing through object and centered around observer is $(4.\pi / 3).v(z)$. Both $\ell_\nu(z)$ and $v(z)$ are specific known functions of $z$ for given cosmological model.

### Calculation of limiting redshift $z_m$

For source of (monochromatic radio) luminosity $L_\nu$, flux density $S_\nu$, (radio) spectral index $\alpha$ ($\equiv$ - dlog $S_\nu$ / dlog $\nu$), and redshift $z$, $L_\nu = 4.\pi.\ell_\nu^2(\alpha, z).S_\nu$. For survey limit $S_0$, value of limiting redshift $z_m$ is given by $\ell_\nu^2(\alpha, z) / \ell_\nu^2(\alpha, z_m) = S_0 / S_\nu \equiv s$, $0 \leq s \leq 1$, for source of redshift $z$ and spectral index $\alpha$. For simplest case,
$[\ell_\nu(\alpha, z) / \ell_\nu(\alpha, z_m)]^2 = s$ has single finite solution $z_m$ for given $\alpha$, $z$ and $S_\nu$, $S_0$. Different values $z_m$ correspond to different $L_\nu(\alpha)$.

### Relating $n(z)$ to $p(V/V_m)$

Let $N(z_m).dz_m$ represent number of sources of limiting redshifts between $z_m$ and $z_m + dz_m$ in sample covering solid angle $\omega$ of sky. Then $4.\pi.N(z_m) / \omega$ is total number of sources of limit $z_m$ per unit $z_m$-interval. Since volume available to source of limit $z_m$ is
$V(z_m) = (4.\pi / 3).(c / H_0)^3.v(z_m)$, (where speed of light $c$ and Hubble constant $H_0$ together determine linear scale of universe,) number of sources (per unit $z_m$-interval) per unit volume is
$\{3.N(z_m) / \omega\}.(H_0 / c)^3.(1 / v_m)$, where $v_m \equiv v(z_m)$. Let $n_m(z_m, z)$ be number of sources / unit volume / unit $z_m$-interval at redshift $z$. Then, $n(z) \equiv \int_z^\infty dz_m. n_m(z_m, z)$, and
$n_m(z_m, z) = \{3.N(z_m) / \omega\}.(H_0 / c)^3.(1 / v(z_m)).p_m(v(z) / v(z_m))$ for $0 \leq z \leq z_m$, where $p_m(x)$ is distribution of $x \equiv V/V_m$ for given $z_m$. For $z > z_m$, $n_m(z_m, z) = 0$, since sources with limiting redshift $z_m$ cannot have $z > z_m$. To get $n(z)$ for all $z_m$-values, integrate over $z_m$:
$n(z) = \{3 / \omega\}.(H_0 / c)^3. \int_z^\infty dz_m.( N(z_m) / v(z_m)).p_m(v(z) / v(z_m))$.

### Scheme of Calculation

Any real sample has maximum $z_{max}$ for $z_m$. So, $n(z_{max}) = 0$. In fact, lifetimes of individual sources will come into consideration, as well as structure-formation epoch at some high redshift (say, > 10). Thus, $n(z)$ calculation will give useful results only upto redshift much less than $z_{max}$. Formally writing $z_{max}$ instead of $\infty$ for upper limit,
$n(z) = \{3 / \omega\}.(H_0 / c)^3. \int_z^{z\_max} dz_m.( N(z_m) / v(z_m)).p_m(v(z) / v(z_m))$ for $0 \leq z \leq z_{max}$.

To apply to real samples, this must be converted to sum. Divide $z_m$-range 0 to $z_{max}$ into k equal intervals, each = $z_{max} / k = \Delta z$. Mid-points are
$z_j = (j - ½).\Delta z = \{(j - ½) / k\}.z_{max}$. Calculate n(z) at these points: $n(z_j)$. Converting integral to sum,
$(\omega / 3).(c / H_0)^3.n(z_j) = \sum_{i=j}^{k} \{N_i / v(z_i)\}.p_i(x_{ij})$, where $x_{ij} = v(zj) / v(z_i)$. **(1)**
It is useful to use $Z_m = \ln z_m$ as redshift variable. Integral and converted sum are then:
$n(z) = \{3 / \omega\}.(H_0 / c)^3. \int_z^{Z\_max} dZ_m.(z_m.N(Z_m) / v(z_m)).p_m(v(z) / v(z_m))$ for $0 \leq z \leq z_{max}$, and
$(\omega / 3).(c / H_0)^3.n(z_j) = \sum_{i=j}^{K} \{z_i.L_i / v(z_i)\}.p_i(x_{ij})$, where $x_{ij} = v(zj) / v(z_i)$. **(1')**
In these two forms (with $z_m$ and $Z_m$ as variables), $N_i$ is population of ith $z_m$-bin and $L_i$ that of ith $Z_m$-bin. There are K bins for $Z_m$, and K and k will, in general, be different.

### Illustrative Calculation in 1981

Wills & Lynds (1978) have defined carefully sample of 76 optically identified quasars. We use this sample only to illustrate derivation of n(z) from $p(x) \equiv p(V/V_m)$. We use Einstein-de Sitter cosmology or $q_0 = \sigma_0 = ½$, $k = \lambda_0 = 0$ or (½, ½, 0, 0) world model in von Hoerner's (1974) notation, for which
$(H_0 / c)^2.\ell_v^2(\alpha, z) = 4.(1 + z)^\alpha / \{\sqrt{(1 + z)} - 1\}^2$ and $(H_0 / c)^3.v(z) = 8.\{1 - 1 / \sqrt{(1 + z)}\}^3$.
For each quasar, $z_m$ is calculated by iteration with initial guess z for $z_m$. Values of z, $z_m$ are then used to calculate v(z), $v(z_m)$ and hence $x = V/V_m$. All 76 $V/V_m$-values are used to plot histogram. Good approximation for p(x) is p(x) = 2.x, which is normalized over [0,1]. The limiting redshifts $z_m$ range from 0 to 3.2. Dividing into four equal intervals, bins centered at 0.4, 1.2, 2.0 and 2.8 contain 19, 31, 16 and 10 quasars. Although each of these 4 subsets is quite small, we calculate and plot histograms $p_i(x)$, i = 1, 2, 3, 4 for each subset for x-intervals of width 0.2 from 0 to 1, thus with 5 intervals centered at x = 0.1, 0.3, 0.5, 0.7 and 0.9. Each normalized $p_i(x)$ is also well approximated by $p_i(x) = 2.x$ except $p_4(0.2994)$. So we do calculations using this approximation in addition to using actual values. Finally we calculate $(\omega / 3).(c / H_0)^3.n(z_j)$ using **(1)** and **(1')**. (See tables.)

**Table for $p_i(x)$ and p(x)**

| x | No. | $p_1(x)$ | No. | $p_2(x)$ | No. | $p_3(x)$ | No. | $P_4(x)$ | No. | p(x) |
|---|---|---|---|---|---|---|---|---|---|---|
| 0.1 | 0 | 0 | 1 | 0.161 | 0 | 0 | 0 | 0 | 1 | 0.066 |
| 0.3 | 2 | 0.526 | 2 | 0.323 | 3 | 0.9375 | 1 | 0.5 | 8 | 0.526 |
| 0.5 | 3 | 0.789 | 6 | 0.968 | 2 | 0.625 | 1 | 0.5 | 12 | 0.789 |
| 0.7 | 8 | 2.105 | 8 | 1.290 | 7 | 2.1875 | 5 | 2.5 | 28 | 1.842 |
| 0.9 | 6 | 1.580 | 14 | 2.258 | 4 | 1.25 | 3 | 1.5 | 27 | 1.776 |
| Totals | 19 | | 31 | | 16 | | 10 | | 76 | |

**Table of n(z) calculation using linear scale for limiting redshifts**

| j | $z_j$ | $N_j$ | → $v(z_j)$ | i = 1 | i = 2 | i = 3 | i = 4 | → $n(z_j)$ |
|---|---|---|---|---|---|---|---|---|
| 1 | 0.4 | 19 | 2.97E-2 | 1 | 0.1074 | 0.0492 | 0.0321 | 1307. |
| 2 | 1.2 | 31 | 0.27666 | | 1 | 0.4580 | 0.2994 | 255. |
| 3 | 2.0 | 16 | 0.60399 | | | 1 | 0.6536 | 67. |
| 4 | 2.8 | 10 | 0.92407 | | | | 1 | 22. |

Notes for second table: (a) $5^{th}$ to $8^{th}$ columns list $x_{ij}$-values,
(b) → $v(z_j) \equiv (H_0 / c)^3.v(z_j) = 8.\{1 - 1 / \sqrt{(1 + z_j)}\}^3$, and
(c) → $n(z_j) \equiv (\omega / 3).(c / H_0)^3.n(z_j)$.

Use of approximations $p_i(x) = 2.x$ in evaluating sums **(1)** for each row j = 1, 2, 3, 4 gives virtually same results. Table below shows steps in evaluating n(z) using ln-scale for limiting redshifts, and $p_i(x) = 2.x$, so that no $x_{ij}$-values need be calculated.

**Table of n(z) calculation using ln-scale for limiting redshifts**

| j | $Z_m$-range | mid-$Z_m$ | $z_m$ (i.e. $z_j$) | $L_j$ | → $v(z_j)$ | → $n(z_j)$ |
|---|---|---|---|---|---|---|
| 1 | -1.5to-0.9 | -1.2 | 0.3012 | 7 | 0.015012 | 355. |
| 2 | -0.9to-0.3 | -0.6 | 0.5488 | 11 | 0.060673 | 301. |
| 3 | -0.3to+0.3 | 0.0 | 1.0000 | 27 | 0.201010 | 337. |
| 4 | +0.3to+0.9 | +0.6 | 1.8221 | 23 | 0.530388 | 181. |
| 5 | +0.9to+1.5 | +1.2 | 3.3201 | 8 | 1.117620 | 48. |

Number of sources in bin j is denoted $L_j$ for ln-scale (instead of $N_j$ for linear scale).

## Conclusion

Due to too small sample, results are only indicative. Main aim is illustrating method fully.

## Acknowledgments

Work reported evolved out of discussions with Vasant K Kulkarni in 1981. Computer Centre of IISc, Bangaluru was used for calculations. First draft was written in 2004-2005 in Muenster, Germany. Radha D Banhatti provided, as always, unstinting material, moral & spiritual support. Uni-Muenster is acknowledged for use of facilities & UGC, New Delhi, India for financial support.

-x0x-